# A low pre-infall mass for the Carina dwarf galaxy from disequilibrium modelling

Uğur Ural[1], Mark I. Wilkinson[2], Justin I. Read[3] & Matthew G. Walker[4]

Dark matter-only simulations of galaxy formation predict many more subhalos around a Milky Way-like galaxy than the number of observed satellites. Proposed solutions require the satellites to inhabit dark matter halos with masses $10^9$–$10^{10}$ Msun at the time they fell into the Milky Way. Here we use a modelling approach, independent of cosmological simulations, to obtain a pre-infall mass of $3.6^{+3.8}_{-2.3} \times 10^8$ Msun for one of the Milky Way's satellites: Carina. This determination of a low halo mass for Carina can be accommodated within the standard model only if galaxy formation becomes stochastic in halos below $\sim 10^{10}$ Msun. Otherwise Carina, the eighth most luminous Milky Way dwarf, would be expected to inhabit a significantly more massive halo. The implication of this is that a population of 'dark dwarfs' should orbit the Milky Way: halos devoid of stars and yet more massive than many of their visible counterparts.

[1] Leibniz Institute für Astrophysik Potsdam, An der Sternwarte 16, Potsdam 14482, Germany. [2] Department of Physics and Astronomy, University of Leicester, University Road, Leicester LE1 7RH, UK. [3] Astrophysics Research Group, Faculty of Engineering and Physical Sciences, University of Surrey, Guildford GU2 7XH, UK. [4] Department of Physics, McWilliams Center for Cosmology, Carnegie Mellon University, 5000 Forbes Avenue, Pittsburgh, Pennsylvania 15213, USA. Correspondence and requests for materials should be addressed to U.U. (email: uural@aip.de).

While the Cold Dark Matter paradigm for structure formation in the Universe has been very successful in reproducing observations on scales larger than ~1 Mpc (refs 1–3), on galactic scales there have been long-standing puzzles. The discrepancy between the predictions of the Cold Dark Matter paradigm and the observed properties of the dwarf spheroidal galaxy satellites (dSphs) of the Milky Way has persisted for over a decade. Numerical models predict that thousands of dark matter subhalos should be found orbiting the Milky Way and Andromeda, yet only a few tens have been found to date[4,5]. This has become known as the 'Missing Satellites' problem. A popular resolution of this issue is to place stars only in the most massive satellite halos, implying a total or 'virial' mass for the Milky Way dSphs of $\sim 10^{10}$ Msun (see Fig. 1). However, in such halos, the required central stellar velocity dispersions of the dSphs would be too high to be consistent with the Milky Way dSphs[6]. More recent work has shown further that more refined mappings between luminous and dark matter result in a population of massive satellites, which have inexplicably failed to form stars (the 'Too Big To Fail' problem)[7–9]. Several solutions have been proposed, including lowering the mass of the Milky Way halo[10], or the central stellar velocity dispersion of the dSphs through the action of stellar feedback[11]. However, these still require the Milky Way dSphs to inhabit dark matter halos with pre-infall masses greater than $\sim 10^{9}$ Msun[4].

Here we use a new 'disequilibrium' model fitting algorithm to constrain the mass of a dwarf galaxy—Carina—that appears to be in the process of tidal disruption by the Milky Way[12]. Previous simulations of Carina, which assumed identical spatial distributions for the dark matter and stars, suggested that tides have played a role in the evolution of this system[13]. The extensive observed data for Carina, combined with our larger parameter space of non-equilibrium models, allows us to measure the mass of Carina over a far greater radial range than has been possible to date. Most significantly, we are able to 'wind the clock' back to estimate its mass before it fell into the Milky Way, without recourse to comparisons with cosmological simulations.

## Results

**Simulation procedure.** Our method works by simulating the disruption of almost 19,000 N-body Carina models in a static Milky Way potential. To marginalize over the unknown model parameters, the N-body models are wrapped up inside a Markov Chain Monte Carlo (MCMC) pipeline (see Methods section). For the dark matter halo, we allowed both cusped and cored central density profiles. The former are found in cold dark matter simulations, while the latter give a better match to observations[14]. The main advantage of the method over previous studies is that by using full N-body simulations, we can model 'disequilibrium' systems like the tidally disrupting Carina dSph. A demonstration of the performance of our methodology on artificial data is shown in Fig. 2: we recover both the pre-infall and present-day mass of a mock dwarf.

**Mass of the Carina dsph.** In Fig. 3, as well as Table 1, we show our results for Carina. Figure 3 shows our best fit surface brightness (top) and projected velocity dispersion (bottom) profiles. Table 1 reports the median and 68% confidence intervals for all of our fitted parameters. To facilitate comparison with the predictions of cosmological simulations, we calculate the distribution of $M_{200}$ (the mass within the radius where the mean density of the dSph reaches 200 times the critical density of the

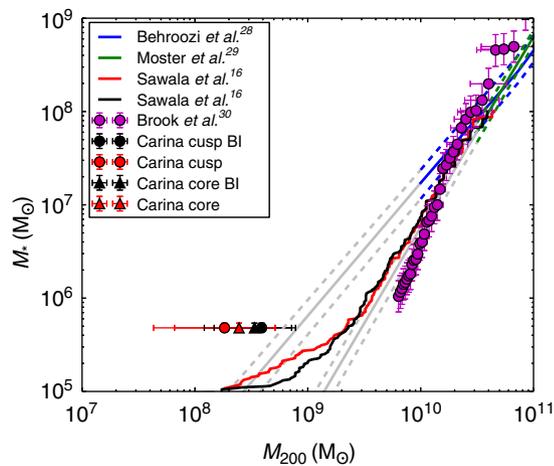

**Figure 1 | The estimated pre-infall mass of the Carina dwarf compared with predictions from cosmological simulations.** The estimated pre-infall mass of the carina dwarf compared with predictions from cosmological simulations. The blue[28] and green[29] lines show abundance matching estimates based on data from the Sloan Digital Sky Survey; below B10^10 Msun (grey lines), they both become extrapolations. The red and black lines show the stellar mass halo mass relation taken from a recent cosmological hydrodynamical simulation[16], using the halo mass before infall (BI: black) and at the present time (red). Unlike the other lines and data points, these two curves are not based on abundance-matching[16]. The purple points show the abundance-matching results between Local Group dwarfs and a 'constrained' simulation of our local volume[30]. The black and red circles are our pre-infall mass estimates ($M_{200}$) for Carina for cusped and cored dark matter halos, respectively. They are lower than any of the curves, only marginally consistent with the extrapolation to low luminosities of the relation found in ref. 28. Finally, the black (cusped) and red triangles (cored) are our present-day mass estimates within 1.5 kpc. We calculate the $1\sigma$ error bars by including 68% of the good models around the median value as given in Table 1.

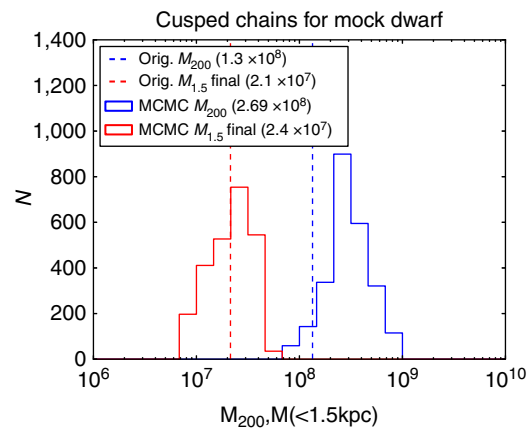

**Figure 2 | The results of MCMC chains run with artificial input data.** The present-day mass is very well constrained (red histogram)—the red dashed line shows the actual value from the target model. The pre-infall mass (blue histogram) has more uncertainty but is still close to the actual value (solid blue line), despite the large range of masses explored by the chains. The values quoted in the legends of the histograms are the median values that the best models in the MCMC chains found. The likely reason for the larger uncertainty in the pre-infall mass is that the target model in this test was chosen to be on a very eccentric orbit and has undergone stronger tidal disturbance than Carina (see Table 1). In addition, the noise we add to its 'observations' (in order to make the uncertainties similar to those in Carina) admits models with a larger range of orbital eccentricities. Both histograms show models with $\chi^2 < 9$.

Universe) values at the start of our simulations. We find pre-infall $M_{200}$ values of $3.9 \times 10^8$ $(-2.4; +3.9)$ and $3.37 \times 10^8$ $(-2.1; +3.8)$ Msun for Carina models with cusped and cored halos, respectively. Interestingly, this low mass estimate agrees with an earlier study, which found that the halos in the Aquarius cosmological simulations that reproduced the mass of Carina at the present time were those with pre-infall masses of $<4 \times 10^8$ Msun[4]. After infall, $M_{200}$ is no longer a meaningful quantity for a satellite galaxy, and we therefore use the mass within 1.5 kpc (the radial extent of our surface brightness and velocity dispersion data) as our present-day mass estimate. This quantity is very well constrained, with $M(r<1.5\,{\rm kpc}) = 7.1 \times 10^7$ $(-3.5; +2.8)$ and $9.7 \times 10^7$ $(-4.8; +4.9)$ Msun for the cusped and cored halos, respectively. We find that both cusped and cored models fit the data very well, suggesting that there is little power in the binned data to distinguish between the two. In general, protected by their higher central densities, cusped halos are able to withstand the stronger tidal forces experienced during closer perigalactic passages.

Our favoured models include both cases in which Carina inhabits a rather low-mass halo, showing significant tidal disruption in its outer parts and cases where it is protected from external tides by a massive halo. These latter models may be favoured by a recent study that found no evidence for tidal tails at a radius of ~1 degree from the main body of Carina[15]. We explicitly tested which of our models show visible tidal tails when analysed similarly to that study and found that only the most tidally disrupted are inconsistent with their results. Table 1 presents results from the three sets of model chains, a cusped and a cored chain that use surface brightness, velocity dispersion and velocity gradient and an additional set of model chains with cusped haloes but which excluded the velocity gradient data. It is seen that ignoring the velocity gradient favours models with an even lower pre-infall mass (nevertheless, consistent with our other results within the errors). We expect our results for Carina to be relatively insensitive to the detailed properties of the Milky Way disk as the majority of successful models are not on disk-crossing orbits.

## Discussion

The most striking aspect of our results is our upper bound on the mass of Carina both today and pre-infall. In Fig. 1, we compare this upper bound to predictions derived from 'abundance matching' schemes, where satellite luminosity is assumed to depend monotonically on the mass of the dark matter subhalos at infall, as well as to cosmological hydrodynamical simulations. Intriguingly, our data point for Carina lies to the low-mass side of all but one of the extrapolated relations. Even in this case, there is a tension at the ~1σ level. Our pre-infall mass estimate, being based on N-body simulations constrained by the observed data for the Carina dSph, does not rely on any assumed cosmological model and, unlike previous studies, it is entirely independent of

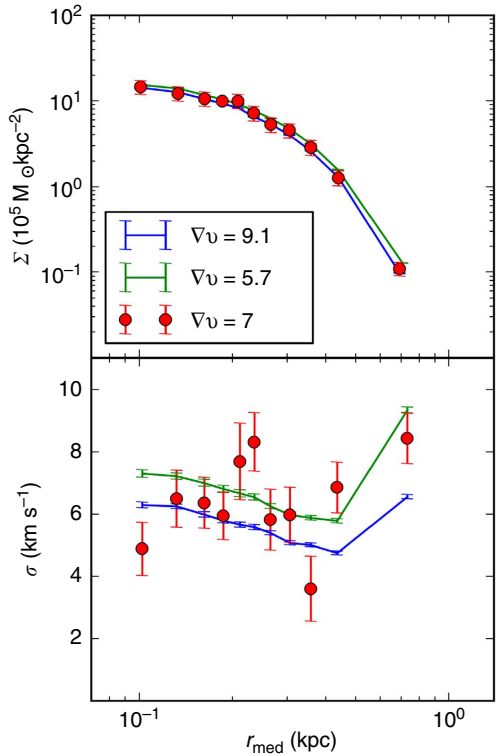

**Figure 3 | A comparison between the best Carina models with cusped and cored dark matter halos and the observed data.** The figure shows the surface brightness and projected velocity dispersion profiles (and the velocity gradient given in km s$^{-1}$ in the legend) from two of our best models. The red data points are the observations with associated 1σ errors. The blue and green curves come from our best fit cusped and cored models, respectively, with their 1σ Poisson errors calculated in each bin.

| Table 1 | Parameter constraints for Carina. | | | |
|---|---|---|---|---|
| | Range | Cusp ($\Sigma$, $\sigma$, $\nabla v$) | Core ($\Sigma$, $\sigma$, $\nabla v$) | Cusp ($\Sigma$, $\sigma$) |
| $M_s$ ($10^5$ Msun; pre-infall) | 4.3; 21.5 | 4.8 (−0.35; +0.6) | 4.8 (−0.35; +0.49) | 4.9 (−0.4; +0.55) |
| $r_s$ (kpc) | 0.074; 0.74 | 0.2 (−0.06; +0.04) | 0.19 (−0.07; +0.06) | 0.22 (−0.07; +0.06) |
| $M_h$ ($10^8$ Msun) | 0.01; 100 | 4.0 (−2.4; +4.1) | 3.48 (−2.25; +3.87) | 2.57 (−1.42; +3.7) |
| $r_h$ (kpc) | $r_s$; 5 | 2.06 (−1.11; +1.96) | 0.88 (−0.46; +0.48) | 1.55 (−0.67; +1.78) |
| $\mu_\alpha \cos(\delta)$ (mas cent$^{-1}$) | −0.17; 0.61 | 0.15 (−0.15; +0.36) | 0.15 (−0.15; +0.33) | 0.23 (−0.21; +0.25) |
| $\mu_\delta$ (mas cent$^{-1}$) | −0.09; 0.57 | 0.11 (−0.11; +0.36) | 0.30 (−0.3; +0.17) | 0.26 (−0.17; +0.18) |
| $R_{\rm peri}$ (kpc) | — | 91.3 (−71.8; +10.7) | 98.6 (−61.9; +3.3) | 98.2 (−62.5; +3.6) |
| $M_{200}$ ($10^8$ Msun; pre-infall) | — | 3.9 (−2.4; +3.9) | 3.37 (−2.1; +3.8) | 2.44 (−1.3; +3.56) |
| $M(r<1.5\,{\rm kpc};\,10^7\,{\rm Msun})$ | — | 7.1 (−3.4; +2.8) | 9.7 (−4.8; +4.9) | 6.1 (−2.4; +3.5) |
| $v_{\rm max,in}$ (pre-infall; km s$^{-1}$) | — | 17.4 (−3.6; +3.2) | 18.3 (−3.6; +5.4) | 15.0 (−2.8; +4.2) |
| $v_{\rm max}$ (present; km s$^{-1}$) | — | 15.0 (−3.1; +2.9) | 17.0 (−3.9; +4.4) | 13.7 (−2.7; +4.0) |

MCMC, Markov Chain Monte Carlo.
The parameters listed in the first column are as follows: stellar mass and scale length ($M_s$, $r_s$); total halo mass we use to generate the model and scale length ($M_h$, $r_h$); the proper motions ($\mu_\alpha \cos(\delta)$, $\mu_\delta$); perigalactic distance ($R_{\rm peri}$); the pre-infall and present epoch masses as explained in the text ($M_{200}$, $M(r<1.5\,{\rm kpc})$); and the maximum halo circular velocity before infall and today ($v_{\rm max,in}$, $v_{\rm max}$). The second column gives the range of values explored by our MCMC chains. Columns three and four list the constraints we obtain for all parameters for chains that used cored or cusped models and included the velocity gradient in the model likelihood. The fifth column presents results for chains with cusped models in which the velocity gradient was ignored. The estimated pre-infall and present-day masses for Carina are calculated as the average values given by our cored and cusped chains.

any cosmological simulations. The low mass that we find thus provides a new and complementary insight into the mapping between low-mass haloes and low-luminosity galaxies and suggests that a simple monotonic mapping between light and dark in our standard cosmological model fails. One solution is to posit that for halo masses below $\sim 10^{10}$ Msun, the physics of galaxy formation leads to a halo occupation function that is effectively stochastic, as suggested by some cosmological hydrodynamic simulations[16,17].

Our findings can be extended in the near future: dwarf galaxy proper motions from the Gaia satellite will significantly decrease the orbital uncertainties for Carina and the other Local Group dwarfs[18]. Combined with deep photometric observations of dwarf outskirts, we will be able to obtain a pre-infall mass distribution for the whole population. This will address the important question of whether Carina is an outlier in the mass distribution. However, even taken in isolation, our determination of the mass of Carina before interacting with the Milky Way precludes any models that associate the most luminous dSphs with the most massive subhalos.

## Methods

**Markov chain Monte Carlo.** We performed a large suite of N-body simulations that compared the final state of the disrupting dwarf spheroidal galaxies in the tidal field of the Milky Way to a host of observational data for the Carina dwarf galaxy. The simulations were performed within a MCMC framework[19] in order to sample the parameter space effectively and constrain the properties of the dSph. Our method is reminiscent of ref. 20 that uses a genetic algorithm combined with restricted N-body simulations to explore the tidal disruption of NGC 205 around the Andromeda galaxy. However, our method is more general, using full rather than restricted N-body simulations and using an MCMC algorithm that allows us to fully explore parameter degeneracies.

First, a two-component, non-rotating and spherical N-body model of the dSph was built with $2 \times 10^5$ particles split equally between the dark matter and stellar components. The starting coordinates and velocities of the dSph were calculated by integrating the orbit of a point mass backwards in a static Milky Way potential using the present-day position and velocity. We then replaced the point mass with the live model and performed a full N-body simulation of its evolution around the Milky Way for 6 Gyr. At the end of the simulation, the radial profiles for the velocity dispersion ($\sigma$) and the surface brightness ($\Sigma$), as well as the velocity gradient ($\nabla v$) between the outermost bins along the major axis of the model dwarf were calculated, and a (reduced) $\chi^2$-statistic was used to compare it with Carina data[12,21]. The likelihood ratio between the consecutive models was used to determine the more favourable region of the parameter space at each step, as the Markov Chain accepted either the new model or re-accepted the older one with the better likelihood. This is the learning process of the MCMC algorithm that chooses new initial conditions for the simulations at each step on the basis of the initial conditions of the last model that is accepted. In our pipeline, the proper motions are used as a prior on the basis of observational data and running N-body simulations that fit both the observational profiles and the proper motions, we are able to marginalize over orbit and halo properties simultaneously with the minimum number of assumptions for the dark matter mass and distribution of the satellite.

Supplementary Fig. 1 shows a schematic representation of the algorithm and the codes used at each step. MCMC chains that used models with cusped and cored halos were run separately, where each N-body simulation took $\sim 1$ h on 32 processors on average. As a very large number of simulations needed to be performed ($\sim 19,000$ in total for Carina), several chains were run in parallel. Before adding the chains together for the final analysis, the first 50–100 simulations of each of them were dismissed to account for the burn-in period of the algorithm.

**Parameter space.** The parameter space of initial conditions consisted of six free parameters ($\mu_\alpha \cos(\delta)$, $\mu_\delta$, $M_s$, $r_s$, $M_h$ and $r_h$) for which the allowed ranges are given in Table 1. Given the potential for systematic errors in the observed proper motions, the allowed range for the present-day proper motions ($\mu_\alpha \cos(\delta)$, $\mu_\delta$) used for the orbit integration was the $3\sigma$ range of the observations (the values of the revised proper motions were obtained through private communication from the authors of ref. 22). The N-body models were generated with falcON[23] according to the split power profiles given in Equation 1, where $\alpha$, $\beta$ and $\gamma$ were fixed: 1,4,1 for the cusped halo; 0,4,1 for the cored halo; and 0.515, 4.45 and 0.287 for the stellar component.

$$\rho = \frac{\rho_0}{\left(\frac{r}{r_s}\right)^\alpha \left(1 + \left(\frac{r}{r_s}\right)^{(\beta-\delta)/\alpha}\right)} \quad (1)$$

The latter was chosen to have a functional form providing a good fit to the present-day surface density in the central regions, albeit with variable amplitude and scale radius that can take values between 0.2 and 2.5 times the original fit ($r_s = 0.237$ kpc). The initial stellar mass $M_s$ was allowed to be up to five times larger than the current one.

The initial mass of the dark matter halo $M_h$ could be as small as $10^6$ Msun, making it only twice as massive as the stellar component, while the upper limit is high enough ($10^{10}$ Msun) to allow the large mass models predicted by the cosmological simulations to be tested. Similarly, while the scale radius of the dark matter halo, $r_h$ can be as large as 5 kpc, the lower limit is determined by that of the stellar component for each model.

**Technical details of the N-body simulations and the Milky Way potential.** The external Milky Way potential (pot 4a[24]) implemented for the point mass orbit integrator[25] and the N-body code PkdGRAV-1 (ref. 26) was provided by the GalPot programme provided in the NEMO Stellar Dynamics Toolbox[27]. The mass of the Milky Way reaches $5 \times 10^{11}$ Msun at 100 kpc, which is Carina's current distance. The simulation time of 6 Gyr was chosen to allow a large enough timescale for the dwarf's evolution while avoiding the complications because of the evolution of the Milky Way potential, and hence Carina's orbit[25].

We keep the parameters describing the potential of the Milky Way fixed in our analysis. This is an additional systematic error that we do not currently marginalize over. However, much of the associated uncertainty is accounted for by the wide range of proper motions admitted by the large observed errors bars and so marginalizing over uncertainties in the Milky Way potential is not yet required. However, as Carina's proper motion errors shrink, this may become the leading error term. Such a situation would open up the possibility of actually constraining the Milky Way potential alongside the mass and orbit of Carina.

The parameters used in PkdGRAV were chosen on the basis of analysis of simulations with different number of particles, opening angle, time step and the softening length, taking into account the trade-off between CPU time and accuracy. Supplementary Table 1 shows a subset of these simulations where it is seen that the main factor that can affect the $\chi^2$ is the number of particles. Therefore, we chose $N = 2 \times 10^5$, the largest number of particles that were computationally affordable—the slow down of the simulation when run with $2 \times 10^6$ particles instead of $2 \times 10^5$ was a factor of 10. We underline that even for the very tidally disrupting model we present in Supplementary Fig. 2, the effects of changing these numerical parameters is not large enough to change the results of the MCMC.

**Tests with mock data.** To test the MCMC pipeline described above, we ran the chains first for a mock dSph instead of Carina. The mock dwarf had a cusped dark matter halo profile to start with and its final $\Sigma$, $\sigma$ and $\nabla v$ were similar to those of Carina. We modified the $\Sigma$ and $\sigma$ profiles, adding Gaussian noise to each data bin as well as increasing the error bars to match the s.d. of the data that resulted in a Mock dwarf with similar uncertainties to Carina. The noise in the velocity gradient in Carina was calculated for $\sim 10$ stars in the outermost data bin on the major axis. The error bars were obtained by using 100,000 random samples of 10 stars. Both the ideal N-body model and the noisy data are presented in Supplementary Fig. 3, where an increase in the velocity dispersion in the outer bins because of the eccentric orbit of the dwarf is observed.

Three sets of Markov Chains were run separately for the mock dwarf. Two chains where the likelihood comparisons between models were made using ($\Sigma$, $\sigma$, $\nabla v$) were run separately testing either cored or cusped dark matter halos, while the third chain with a cusped halo used only $\Sigma$ and $\sigma$ in its 'observations'. Supplementary Table 2 shows the parameter values used in the mock data as well as those found by the three chains. The $\chi^2$ cuts were chosen to include as many as possible 'good' models for our statistical calculations to be meaningful. The numbers presented in the Supplementary Table 2 are derived from 2,469 (50 unique) models with $\chi^2$ ($\Sigma$, $\sigma$, $\nabla v$) $<9$ and 3,119 (335 unique) models with $\chi^2$ ($\Sigma$, $\sigma$) $<6$ for the cusped chains.

It is seen from Supplementary Table 2 that the current mass $M(r<1.5$ kpc) is very well constrained by both the cored and the cusped chains within the radius probed by the data. It is also seen that when the velocity gradient was excluded from the $\chi^2$ calculations, the chains found models with larger and more massive dark matter halos and hence overestimated the both the initial and the final mass. Interestingly, although the Mock dwarf was cusped to start with, the cored chains recovered the initial mass better than those with cusps that found models with M200 twice as massive. Nevertheless, the results for these chains are very promising considering the wide ranges of masses the algorithm explored. Supplementary Fig. 4 and Supplementary Fig. 5 show the distribution of the good models in the parameter space that was explored by the cusped (cored) chains, for all of the models as well as those within the $\chi^2<9$ cut. The figure shows that the orbit is the worst constrained property of the models; hence, any improvement of the observed uncertainties in the proper motions will improve the results.

As seen in Fig. 2, the final mass is very well constrained by the chains within the radius probed by the data.

**MCMC for Carina.** The same three sets of MCMC chains were then run for Carina, which is ideally suited for our study because of the quality of available photometric and spectroscopic data extending to large radii, as well as the evidence of a velocity gradient along the major axis[12]. These new chains found more models with small $\chi^2$ than those for the mock dwarf. In order to test the validity of the $\chi^2$ cut used above, we re-analysed the Carina chains for different cuts. Supplementary Figs 6 and 7 show the distribution of the models for a $\chi^2 < 6$ cut, which included 445 models (22 unique) and the $\chi^2 < 9$ cut that we use throughout this paper with 2,686 models (232 unique). As can be seen from the figures, these two options result in mass estimates consistent with each other. Therefore, as the overall power of the method we use is to provide a distribution of good models, we choose to use the same $\chi^2$ cut for Carina and the mock dwarf. As a final test, in Supplementary Fig. 8, we compare two models with different $\chi^2$ for both cusped and cored halos and demonstrate that even at the highest end of the $\chi^2$ range the models used still provide a reasonable match for Carina.

**Code availability.** The simulation data, MCMC algorithm, pytipsy package necessary to read the binary data files and the codes analysing the simulation data are available on the project website: http://vo.aip.de/dwarfedmasses/.

The new version of PkdGRAV that was used for the N-body simulation can be obtained from https://hpcforge.org/projects/pkdgrav2/.

The falcON code used to generate the two-component N-body models is part of the NEMO package available at https://github.com/Milkyway-at-home/nemo/tree/master/nemo_cvs/usr/dehnen/falcON.

## Acknowledgements

We thank the developers of the codes we used in the pipeline: Hanni Lux for her orbit integration code; Walter Dehnen for the GalPot and falcON codes that we used to generate our two-component galaxies; Joachim Stadel for PkdGRAV; Jonathan Coles for his tipsy routines. We also thank Kristin Riebe who prepared the website of the project as well as the cover art. U.U. acknowledges funding for a large part of this project from the European Commission under the Marie Curie Host Fellowship for Early Stage Research Training SPARTAN, Contract No MEST-CT-2004-007512, University of Leicester, UK. M.I.W. acknowledges the Royal Society for support via a University Research Fellowship. J.I.R. would like to acknowledge support from SNF grant PP00P2 128540/1. M.G.W. is supported by National Science Foundation grants AST-1313045 and AST-1412999. The simulations in this paper used the Complexity HPC cluster at the University of Leicester, which is part of the DiRAC2 national facility, jointly funded by STFC and the Large Facilities Capital Fund of BIS.

## Author contributions

U.U. performed and analysed all the simulations described in this paper. M.G.W. calculated the velocity dispersion and mean velocity profiles on the basis of his latest observed data on the Carina dSph. M.I.W. and J.I.R. assisted with technical aspects of the numerical simulations and provided input on the cosmological implications of the work. All authors contributed to the writing of the paper.

## Additional information

**Supplementary Information** accompanies this paper at http://www.nature.com/naturecommunications

**Competing financial interests:** The authors declare no competing financial interests.

**Reprints and permission** information is available online at http://npg.nature.com/reprintsandpermissions/

**How to cite this article:** Ural, U. *et al.* A low pre-infall mass for the Carina dwarf galaxy from disequilibrium modelling. *Nat. Commun.* 6:7599 doi: 10.1038/ncomms8599 (2015).